\documentclass[twocolumn,showpacs,preprintnumbers,aps,prb]{revtex4}

\newcommand{\BE}{\begin{equation}}
\newcommand{\EE}{\end{equation}}
\newcommand{\BA}{\begin{eqnarray}}
\newcommand{\EA}{\end{eqnarray}}
\usepackage{graphicx}% Include figure files
\usepackage{dcolumn}% Align table columns on decimal point
\usepackage{bm}% bold math

\begin{document}

\preprint{0201}

\title{Localization of electromagnetic waves in a two dimensional random medium}

\author{Zhen Ye}
\altaffiliation[Corresponding author: ]{Department of Physics,
National Central University, Taiwan} \email{zhen@phy.ncu.edu.tw}
\author{Sheng Li}
\author{Xin Sun}
\affiliation{Research Center for Theoretical Physics, Fudan
University \\
and National Laboratory of Infrared Physics, Shanghai, China}

%\date{February 2, 2002}
\date{\today}

\begin{abstract}

Motivated by previous investigations on the radiative effects of
the electric dipoles embedded in structured cavities, localization
of electromagnetic waves in two dimensions is studied {\it ab
initio} for a system consisting of many randomly distributed two
dimensional dipoles.
%(e.~g. T. Erdogan et al., J. Opt. Soc. Am. B 10, 391 (1993))
A set of self-consistent equations, incorporating all orders of
multiple scattering of the electromagnetic waves, is derived from
first principles and then solved numerically for the total
electromagnetic field. The results show that spatially localized
electromagnetic waves are possible in such a simple but realistic
disordered system. When localization occurs, a coherent behavior
appears and is revealed as a unique property differentiating
localization from either the residual absorption or the
attenuation effects.

\end{abstract}

\pacs{42.25.Hz, 41.90.+e} \maketitle

When propagating through a medium consisting of many scatterers,
waves will be scattered by each scatterer. The scattered waves
will be again scattered by other scatterers. Such a process will
be repeated to establish an infinite recursive pattern of multiple
scattering. As a result, the wave propagation may be significantly
altered\cite{Ishimaru}. It is now well-known that the multiple
scattering of waves is responsible for many fascinating phenomena,
ranging from phenomena of macroscopic scales such as twinkling
lights in the evening sky, modulation of ambient noise in the
oceans, and electromagnetic scintillation of turbulence in the
atmosphere, to phenomena of microscopic or mesoscopic scales
including random lasers\cite{rl} and electronic resistivity in
disordered solids\cite{er}. It has also been proposed that under
certain conditions, the multiple scattering can lead to the
unusual phenomenon of wave localization, a concept originally
introduced to describe disorder induced metal-insulator
transitions in electronic systems\cite{Anderson}.

Over the past two decades, localization of classical waves has
been under intensive investigations, leading to a very large body
of literature(e.~g.
\cite{HW,John,Kirk1,He,Condat,Sor,Genack,McCall,McCall2,Ad,Marian,Sigalas,AAA,Sheng,Wiersma,weak,AAC}).
%and monographs such as\cite{Sheng}.
Such a localization phenomenon has been characterized by two
levels. One is the weak localization associated with the enhanced
backscattering. That is, waves which propagate in the two opposite
directions along a loop will obtain the same phase and interfere
constructively at the emission site, thus enhancing the
backscattering. The second is the strong localization, without
confusion often just termed as localization, in which a
significant inhibition of transmission appears and the energy is
mostly confined spatially in the vicinity of the emission site.

While the weak localization, regarded as a precursor to the strong
localization, has been well studied both theoretically (e.~g. the
monograph\cite{er,Sheng}) and experimentally (e.~g. \cite{weak}),
observation of strong localization of classical waves for higher
than one dimension remains a subject of
debate\cite{Wiersma,AAC,debate}, primarily because a suitable
system is hard to find and the observation is often obscured by
such effects as the residual absorption\cite{debate} and
scattering attenuation\cite{Ye3}.

In this paper, we wish to present a simple but realistic system to
study the phenomenon of strong localization in two dimensions. The
system we consider stems from the previous research on enhancement
and inhibition of electromagnetic radiation in structured media.
For example, Kuhn\cite{Kuhn} used the model to describe the effect
of a metallic mirror on the radiation from a nearby excited
molecule, whereas Chance et al.\cite{Chance} used to explain the
experimental data of Drexhage. Later, Erdogan et al.\cite{Erdogan}
and Wang et al.\cite{ccw} employed the model to analyze the
effects of the cylindrically periodic structure on the radiation
of an enclosed two-dimensional (2D) dipole. Inspired by the work
of Erdogan et al.\cite{Erdogan}, we attempt to construct a system
consisting of many 2D dipoles. The propagation of electromagnetic
(EM) waves in such a system is formulated rigorously in terms of a
set of coupled equations, and then is solved numerically. We show
that a strong localization of EM waves is possible in this system.
In line with the work on acoustic localization\cite{Ye2}, a phase
diagram method is used to describe localization of EM waves. Since
the system considered here results from the practical application
of light emission in structures, an experimental verification may
be readily realized.

Following Erdogan et al.\cite{Erdogan}, we consider 2D dipoles as
an ensemble of harmonically bound charge elements. In this way,
each 2D dipole is regarded as a single dipole line, characterized
by the charge and dipole moment per unit length. Assume that $N$
parallel dipole lines, aligned along the $z$-axis, are embedded in
a uniform dielectric medium and {\it randomly} located at
$\vec{r}_i (i=1,2,\dots,N)$. The averaged distance between dipoles
is $d$. A stimulating dipole line source is located at
$\vec{r}_s$, transmitting a continuous wave of angular frequency
$\omega$. By the geometrical symmetry of the system, we only need
to consider the $z$ component of the electrical waves.

Upon stimulation, each dipole will radiate EM waves. The radiated
waves will then repeatedly interact with the dipoles, forming a
process of multiple scattering. The equation of motion for the
$i$-th dipole is \BA \frac{d^2}{dt^2}p_i + \omega_{0,i}^2p_i &=&
\frac{q_i^2}{m_i}E_{z}(\vec{r}_i) - b_{0,i}\frac{d}{dt}
p_i,\nonumber\\
& & \mbox{for} \ i = 1, 2,\dots, N. \label{eq:1}\EA where
$\omega_{0,i}$ is the resonance (natural) frequency, $p_i$, $q_i$
and $m_i$ the dipole moment, charge and effective mass per unit
length of the $i$-th dipole respectively. $E_{z}({\vec{r}_i})$ is
the total electrical field acting on dipole $p_i$, which includes
the radiated field from other dipoles and also the directly field
from the source. The factor $b_{0,i}$ denotes the damping and is
determined by energy conservation. Without dissipation, $b_{0,i}$
can be determined from the balance between the radiative and
vibrational energies, and is given as\cite{Erdogan} \BE b_{0,i} =
\frac{q_i^2\omega_{0,i}}{4\epsilon m_i c^2}, \label{eq:2}\EE with
$\epsilon$ being the constant permittivity and $c$ the speed of
light in the medium separately.

Equation (\ref{eq:1}) is similar to Eq.~(1) in \cite{Erdogan}. The
difference lies in the driving field on the right hand side of the
equation. In \cite{Erdogan}, $E_z$ is the reflected field at the
dipole due to the presence of reflecting surrounding structures,
while in the present case the field is from the stimulating source
and the radiation from all other dipoles.

The transmitted electrical field from the continuous line source
is determined by the Maxwell equations\cite{Erdogan}\BE \left
(\nabla^2 - \frac{\partial^2}{c^2\partial t^2}\right
)G_0(\vec{r}-\vec{r}_s) =
-4\mu_0\omega^2p_0\pi\delta(\vec{r}-\vec{r}_s) e^{-i\omega t},
\label{eq:3}\EE where $\omega$ is the radiation frequency, and
$p_0$ is the source strength and is set to be unit. The solution
of Eq.~(\ref{eq:3}) is clearly \BE G_0(\vec{r}-\vec{r_s}) =
(\mu_0\omega^2p_0) i\pi H_0^{(1)}(k|\vec{r}-\vec{r}_s|)
e^{-i\omega t}, \label{eq:4}\EE with $k=\omega/c$, and $H_0^{(1)}$
being the zero-th order Hankel function of the first kind.

Similarly, the radiated field from the $i$-th dipole is given by
\BE \left (\nabla^2 - \frac{\partial^2}{c^2\partial t^2}\right
)G_i(\vec{r}-\vec{r}_i) =
\mu_0\frac{d^2}{dt^2}p_i\delta^{(2)}(\vec{r}-\vec{r}_i).
\label{eq:5}\EE The field arriving at the $i$-th dipole is
composed of the direct field from the source and the radiation
from all other dipoles, and thus is given as \BE E_z(\vec{r}_i) =
G_0(\vec{r}_i - \vec{r}_s) + \sum_{j=1, j\neq i}^N G_j(\vec{r}_i -
\vec{r}_j). \label{eq:6}\EE

Substituting Eqs.~(\ref{eq:4}), (\ref{eq:5}), and (\ref{eq:6})
into Eq.~(\ref{eq:1}), and writing $p_i = p_ie^{-i\omega t}$, we
arrive at \BA (-\omega^2 + \omega^2_{0,i} - i\omega b_{0,i})p_i =&
&\nonumber\\ \frac{q_i^2}{m_i}\left[G_0(\vec{r}_i - \vec{r}_s) +
\sum_{j=1, j\neq i}^N \frac{\mu_0 \omega^2}{4\pi}
iH_0^{(1)}(k|\vec{r}_i-\vec{r}_j|)p_i\right].& &  \label{eq:7}\EA
This set of linear equations can be solved numerically for $p_i$.
After $p_i$ are obtained, the total field at any space point can
be readily calculated from \BE E_z(\vec{r}) = G_0(\vec{r} -
\vec{r}_s) + \sum_{j=1}^N G_j(\vec{r} - \vec{r}_j).\label{eq:8}\EE

In the standard approach to wave localization, waves are said to
be localized when the square modulus of the field
$|E(\vec{r})|^2$, representing the wave energy, is spatially
localized after the trivial cylindrically spreading effect is
eliminated. Obviously, this is equivalent to say that the further
away is the dipole from the source, the smaller its oscillation
amplitude, expected to follow an exponentially decreasing pattern.

To this end, it is instructive to point out that an alternative
two dimensional dipole model was deviced previously by Rusek and
Orlowski\cite{Marian}. The authors derived a set of linear
algebraic equations, which is similar in form to the above
Eq.~(\ref{eq:7}). However, there are some fundamental differences
between the two models. In \cite{Marian}, the interaction between
dipoles and the external field is derived by the energy
conservation, while in the present case the coupling is determined
without ambiguity by the Newton's second law. The former leads to
an undetermined phase factor. According to, e.~g.
Refs.~\cite{Erdogan,ccw}, the energy conservation can only give
the radiation factor in Eq.~(\ref{eq:2}). We would also like to
point out that the set of couple equations in Eq.~(\ref{eq:7}) is
similar in spirit to the tight-binding model used to study the
electronic localization\cite{Anderson,Zallen}.

There are several ways to introduce randomness to
Eq.~(\ref{eq:7}). For example, the disorder may be introduced by
randomly varying such properties of individual dipoles as the
charge, the mass or the two combined. This is the most common way
that the disorder is introduced into the tight-binding model for
electronic waves\cite{Zallen}. In the present study, the disorder
is brought in by the random distribution of the dipoles.

Before moving to solve Eq.~(\ref{eq:7}) for the phenomenon of
localization of EM waves, we discuss a general property of wave
localization. The energy flow of EM waves is $\vec{J} \sim
\vec{E}\times\vec{H}$. By invoking the Maxwell equations to relate
the electrical and magnetic fields, we can derive that the time
averaged energy flow is \BE <\vec{J}>_t \equiv \frac{1}{T}\int_0^T
dt \vec{J} \sim |\vec{E}|^2\nabla\theta, \label{eq:9}\EE where the
electrical field is written as $\vec{E} = \vec{e}_E
|\vec{E}|e^{i\theta}$, with $\vec{e}_E$ denoting the direction,
$|\vec{E}|$ and $\theta$ being the amplitude and the phase
respectively. It is clear from Eq.~(\ref{eq:9}) that when $\theta$
is constant, at least by spatial domains, while $|\vec{E}| \neq
0$, the flow would come to a stop and the energy will be localized
or stored in the space. In the localized state, a source can no
longer radiate energies. Alternatively, we can write the
oscillation of the dipoles as $p_i = |p_i|e^{i\theta_i}$. By
studying the square modulus of $p_i$ in the form of
$|\vec{r}_i-\vec{r}_s||p_i|^2$, and its phase $\theta_i$, we can
also investigate the localization of EM waves. Note here that the
factor $|\vec{r}_i-\vec{r}_s|$ is to eliminate the cylindrical
spreading effect in 2D as can be seen from the expansion of the
Hankel function $|H_0^{(1)}(x)|^2 \sim \frac{1}{x}$. That the
phase $\theta$ is constant implies that a coherence behavior
appears in the system. In other words, the localized state is a
phase-coherent state. This picture of localization also holds for
acoustic and electronic waves. It is a unique feature of wave
localization.

For simplicity yet without losing generality, assume that all the
dipoles are identical and they are randomly distributed within a
square area. The source is located at the center (set to be the
origin) of this area. For convenience, we make Eq.~(\ref{eq:7})
non-dimensional by scaling the frequency by the resonance
frequency of a single dipole $\omega_0$. This will lead to two
independent non-dimensional parameters $b = \frac{q^2\mu_0}{4m}$
and $b_0^\prime =
\frac{\omega}{\omega_0}\left(\frac{b_0}{\omega_0}\right)$. Both
parameters may be adjusted in the experiment. For example, the
factor $b_0$ can be modified by coating layered structures around
the dipoles\cite{Erdogan}. Then Eq.~(\ref{eq:7}) becomes simply
\BA (-f^2+1 -ib_0^\prime)p_i =
& & \nonumber\\
ibf^2\left[p_0H_0^{(1)}(k|\vec{r}_i-\vec{r}_s|) + \sum_{j=1, j\neq
i}^N p_iH_0^{(1)}(k|\vec{r}_i-\vec{r}_j|)\right] && \label{eq:10})
\EA with $f=\frac{\omega}{\omega_0}$. Eq.~(\ref{eq:10}) can be
solved numerically for $p_i$ and then the total field is obtained
through Eqs.~(\ref{eq:3}), (\ref{eq:5}) and (\ref{eq:8}). In the
calculation, we scale all lengths by the averaged distance between
dipoles $d$. In this way, the frequency $\omega$ always enters as
$kd$ and the natural frequency $\omega_0$ as $k_0d$. We find that
all the results shown below are only dependent on parameters $b$,
$b_0/\omega_0$, and the ratio $\omega/\omega_0$ or equivalently
$k/k_0 = (kd)/(k_0d)$. Such a simple scaling property may
facilitate designing experiments.

We have first computed the transmitted intensity averaged over the
random distribution of the dipoles as a function of
non-dimensional frequency $kd$. The results indicate that the
natural frequency of the dipoles is altered by the multiple
scattering of EM waves, and is shifted towards the lower frequency
end. We also find that the transmission is significantly
suppressed in a range of frequencies slightly above the natural
frequency of a single dipole, suggesting the strong localization
in this range of frequencies. This helps us to search for the
frequencies where the strong localization is most prominent. We
show an example below.

\input epsf.tex
%\begin{center}
\begin{figure}
\vspace{10pt} \epsfxsize=2.5in\epsffile{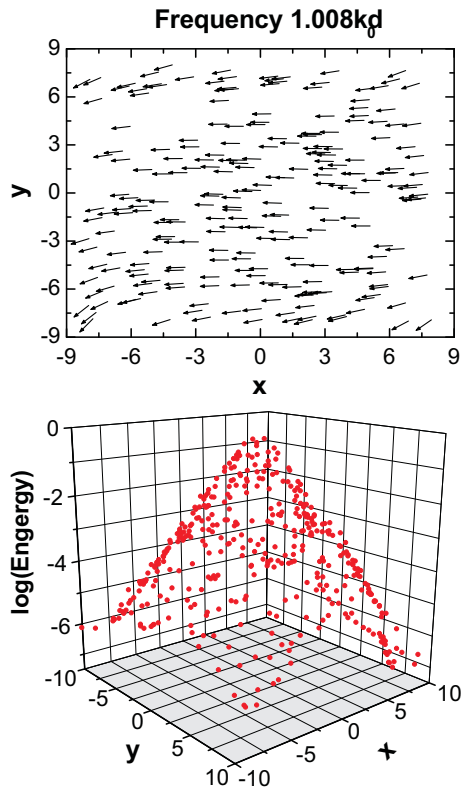} \caption{Top:
The phase diagram for the two dimensional phase vectors defined in
the text; the phase of the source is assumed to be zero. Bottom:
The spatial distribution of energy ($\sim |E|^2$). Here the
geometrical factor has been removed.} \label{fig1}
\end{figure}
%\end{center}

By setting $b = b_0/\omega_0 = 10^{-3}$ and $k_0d=1$, we find a
strong localization of EM waves at, for instance, $\omega/\omega_0
= 1.008$. In this case, the natural frequency of the dipoles is
shifted to around 0.9$\omega_0$. A typical picture of wave
localization is shown in the bottom part of Fig.~\ref{fig1} for an
arbitrary random distribution of 400 dipoles. To describe the
phase behavior of the system, we assign a unit phase vector,
$\vec{u}_i = \cos\theta_i\vec{e}_x + \sin\theta_i\vec{e}_y$ to the
oscillation phase $\theta_i$ of the dipoles. Here $\vec{e}_x$ and
$\vec{e}_y$ are unit vectors in the $x$ and $y$ directions
respectively. These phase vectors are represented by a phase
diagram in the $x-y$ plane with the phase vector $\vec{u}_i$ being
located at the dipole to which the phase $\theta_i$ is associated.
The phase behavior of the system is depicted by the top portion of
Fig.~\ref{fig1}.

Here it is clearly shown that the field energy is strongly
localized near the transmitting source, and, as expected, the
energy decreases almost exponentially along the radial direction.
Meanwhile, the system reveals an in-phase phenomenon: nearly all
the phase vectors of the dipoles point to the same direction,
exactly opposite to the phase vector of the source, i.~e. the
dipoles tend to oscillate out of phase with the source. It is easy
to see that such a coherent behavior effectively prevents waves
from propagation. The picture represented by Fig.~\ref{fig1} fully
complies with the general description of wave localization stated
above. The energy localization in Fig.~\ref{fig1} is also in
qualitative agreement with that observed in the microwave
localizaton\cite{McCall,McCall2} and the acoustic localization in
water with air-cylinders\cite{Ye2}.

We note from Fig.~\ref{fig1} that near the sample boundary, the
phase vectors start to point to different directions. This is
because the numerical simulation is carried out for a finite
sample size. For a finite system, the energy can leak out at the
boundary, resulting in disappearance of the phase
coherence\cite{Wang}. When enlarging the sample size by adding
more dipoles while keeping the averaged distance between dipoles
fixed, we observe that the area showing the perfect phase
coherence will increase accordingly. Another factor affecting the
phase coherence behavior, and subsequently the wave localization,
is the damping factor. When the absorption is added, the in-phase
behavior will be degraded, and the waves will become delocalized
gradually, in agreement with the previous simulation on acoustic
localization in bubbly water\cite{Wang}. Though degrading, we find
that the localization can sustain a substantial variation in the
damping factor, making the system a good candidate for observing
the localization phenomenon.

\input epsf.tex
\begin{figure}
\epsfxsize=2.5in\epsffile{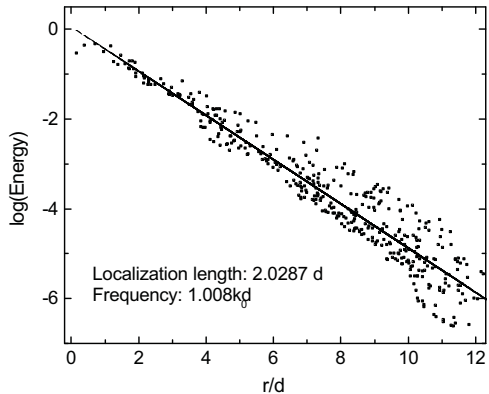} \caption{Energy versus the
distance away from the source. Again we removed the cylindrical
spreading factor.} \label{fig2}
\end{figure}

To find the localization length in Fig.~\ref{fig1}, we plot the
total energy in all directions as a function of the distance from
the source. The results are presented by Fig.~\ref{fig2}. Here,
the numerical data are shown by the black squares, and the result
fitted from the least squares method is shown by the solid line.
It shows that after removing the spreading factor, the data can be
fitted by $e^{-r/\xi}$. From the slop of the solid line, the
localization length $\xi$ is estimated as around $2.02d$, which is
in the vicinity of the localization length $1.6d$ estimated
experimentally for microwave localization in 2D dielectric
lattices\cite{McCall2}.

In summary, we have presented a model system to study the
localization of EM waves in 2D random media. The results show that
spatially localized EM waves can be realized in this simple but
realistic disordered system. When localization occurs, a coherent
behavior appears as a unique feature separating the localization
from other effects such as the residual absorption or extinction
effects. Such a coherence phenomenon for waves in random media
could also be relevant to the random laser action\cite{rl}.

This work is supported by the Bai Yu Lan Fund of Shanghai and the
NSF of China (20074007, 90103034).

\end{document}